\documentclass[modern]{aastex631}
\usepackage[utf8]{inputenc}
\usepackage{amsmath}
\usepackage{MnSymbol}
\usepackage{rotating}

\renewcommand{\twocolumngrid}{}
\renewcommand{\paragraph}[1]{\medskip\par\noindent\textbf{#1}~---}

\usepackage{graphicx}
\usepackage{xcolor}
\usepackage[framemethod=tikz]{mdframed}
\usetikzlibrary{shadows}
\definecolor{captiongray}{HTML}{555555}
\mdfsetup{%
innertopmargin=2ex,
innerbottommargin=1.8ex,
linecolor=captiongray,
linewidth=0.5pt,
roundcorner=1pt,
shadow=false,
}
\newlength{\figurewidth}
\shorttitle{Constrained linear models for normalising stellar spectra}
\shortauthors{Casey et al.}

\newcommand{\project}[1]{\textit{#1}}
\renewcommand{\vec}[1]{\mathbf{#1}}

\newcommand{\vecalpha}{\boldsymbol{\alpha}}
\newcommand{\vecbeta}{\boldsymbol{\beta}}
\newcommand{\vecgamma}{\boldsymbol{\gamma}}
\newcommand{\vecW}{\mathbf{W}} 
\newcommand{\vecG}{\mathbf{G}} 
\newcommand{\vecH}{\mathbf{H}} 

\newcommand{\hadamard}{\odot}

\newcommand{\eso}{\project{ESO}}
\newcommand{\harps}{\project{HARPS}}

\newcommand*{\transpose}{^{\mkern-1.5mu\mathsf{T}}}

\definecolor{tab:blue}{HTML}{1170aa}
\definecolor{tab:red}{HTML}{d1615d}

\usepackage[normalem]{ulem} 

\renewcommand{\added}[1]{#1}
\newcommand{\removed}[1]{}

\sloppypar
\begin{document}

\title{A constrained linear model for continuum normalization of stellar spectra}

\author[0000-0003-0174-0564]{Andrew R. Casey}
\affiliation{School of Physics \& Astronomy, Monash University, Australia}
\affiliation{Centre of Excellence for Astrophysics in Three Dimensions (ASTRO-3D)}
\affiliation{Center for Computational Astrophysics, Flatiron Institute, 162 Fifth Avenue, New York, NY 10010, USA}

\author[0000-0001-7339-5136]{Adam Wheeler}
\affiliation{Center for Computational Astrophysics, Flatiron Institute, 162 Fifth Avenue, New York, NY 10010, USA}

\author[0000-0001-9907-7742]{Megan Bedell}
\affiliation{Center for Computational Astrophysics, Flatiron Institute, 162 Fifth Avenue, New York, NY 10010, USA}

\author[0000-0003-2866-9403]{David W. Hogg}
\affiliation{Center for Cosmology and Particle Physics, Department of Physics, New York University}
\affiliation{Max-Planck-Institut f\"ur Astronomie, Heidelberg}
\affiliation{Center for Computational Astrophysics, Flatiron Institute, 162 Fifth Avenue, New York, NY 10010, USA}

\author[0000-0002-6561-9002]{Andrew Sayjdari}
\affiliation{Department of Astrophysical Sciences, Princeton University, Princeton, NJ 08544, USA}
\affiliation{Department of Physics, Harvard University, 17 Oxford St., Cambridge, MA 02138, USA}
\affiliation{Harvard-Smithsonian Center for Astrophysics, 60 Garden St., Cambridge, MA 02138, USA}

\author[0000-0002-3852-3590]{Lily Zhao}
\affiliation{Department of Astronomy \& Astrophysics, University of Chicago, 5640 S Ellis Avenue, Chicago, IL 60637, USA}
\affiliation{Center for Computational Astrophysics, Flatiron Institute, 162 Fifth Avenue, New York, NY 10010, USA}

\begin{abstract}\noindent
Inferring stellar parameters and chemical abundances by forward modelling stellar spectra usually requires a spectral synthesis code, or an emulator constructed from a curated training set. In these situations continuum normalization is often implemented as a pre-processing step that is independent of stellar parameters. This leads to results that are biased, or inconsistent across signal-to-noise ratios. A more justified approach is to forward model spectra with all nuisances simultaneously, but in practice this can be an expensive or non-convex optimisation procedure. Here we describe a constrained \emph{linear} model that can fit stellar absorption, telluric transmission, the joint continuum-instrument response. Stellar absorption and telluric transmission are each modelled by factorising a grid of rectified theoretical spectra into two non-negative matrices with a chosen number of basis components. This model characterises all possible spectra in many fewer parameters than comparable data-driven models. The non-negativity constraint ensures basis vectors are strictly additive, which limits rectified flux to less than or equal to unity, such that we can distinguish normalized spectra from the joint instrument-continuum response. The model requires no initial guess, and the linearity ensures that inference is convex, stable, and fast. This model allows us to reliably fit nuisances (e.g., tellurics, continuum), and is readily extensible to radial velocity and rotational broadening, without any prior knowledge about the fundamental stellar properties. We demonstrate our method by fitting \eso/\harps\ high-resolution echelle spectra of BAFGKM-type stars. With repeat observations of $\alpha$-Centauri A we present results that are best in class: consistent across time to 0.2\% at S/N $\sim$ 100, and to better than 0.5\% at S/N $\sim$ 30.\\
\end{abstract}


\section{Introduction}\label{sec:intro}

Continuum normalization\footnote{Or continuum rectification, as it was historically described.} is often required before estimating stellar parameters and chemical abundances. Those quantities are inferred from line strengths that are measured \emph{relative} to the continuum, in part because modeling the stellar continuum (and instrument response) correctly is more challenging than modeling the relative flux \citep[e.g.,][]{Wheeler:2023, Wheeler:2024}.
The practice of continuum rectification is ripe with subjectivity, in part for justified historical reasons. Often a consistent procedure is preferable, so people tend to stick with what they know, and that subjectivity is inherited by graduate students. While there is general agreement in the literature that is important to achieve consistent continuum normalization \citep{Blanco-Cuaresma:2014}, there is no apparent consensus on how it should be done.\\

The literature describes a variety of rectification strategies used to achieve different goals \citep{Ness:2015, Casey:2016, Xu:2019, Cretignier:2020}. \added{In detailed studies of individual objects (e.g., to derive accurate atmospheric parameters or chemical abundances from equivalent widths) a} \removed{Classical} spectroscopist might make some coarse estimate of the continuum for the spectrum, and locally refine that estimate for every absorption line of interest \citep[e.g.,][]{Casey:2014}. This process is often done by hand,\footnote{Albeit they are often experienced hands. See \citet{Bensby:2014}: `\emph{more than 300\,000 equivalent widths were measured by (the first author's right) hand}'.} although some graphical user interface tools exist to streamline the process and improve reproducibility. This approach usually restricts the spectroscopist to only measuring strengths of what are believed to be isolated lines: atomic transitions that are not contaminated by neighbouring lines or molecular bands. In contrast, \added{large spectroscopic surveys require automatic extraction of thousands of normalized spectra, where the primary objective is a} \removed{industrial spectroscopists often have a different objective: one where they solely for a} \emph{consistent} continuum normalization procedure \citep[e.g.,][]{Ness:2015, Casey:2016}. In those situations the process does not have claim to deliver the true continuum, but it should deliver a \emph{pseudo}-continuum that is consistent for stars of similar stellar parameters and signal-to-noise (S/N) ratios.\\

Fitting the continuum correctly requires you to know where there is absorption. But to know where to expect line absorption, we must have some estimate of the stellar parameters (among other things, e.g., line lists). Without knowing the stellar parameters -- or having a good model for line absorption -- spectroscopists have developed various bespoke methods. A popular choice is to iteratively mask pixels in an asymmetric manner (so-called `sigma clipping') to exclude data points some chosen level below the current estimate of the continuum. This works well if the practictioner can confidently set the bounds on where to exclude data. However, those parameters would need to be different for a more metal-rich star, or for a spectrum with a low S/N ratio. Exactly how to specify the asymmetric clipping parameters comes down to experience.\\

An alternative is simply to mask all pixels except a carefully selected set of so-called `continuum pixels'. This is painstaking work to perform manually, but can be done by iteratively training data-driven models \citep{Ness:2015}, or building masks from approximate line strengths computed from atomic properties. In either scenario, the set of continuum pixels is only valid for stars of a similar type and metallicity. In many cases there are simply \emph{no} continuum pixels (e.g., near a molecular absorption band in an M-dwarf). Some graphical tools exist to enable practictioners to select pseudo-continuum points to `anchor' some continuum fit \citep{Xu:2019,Cretignier:2020} which are useful if the practictioner wants full control over the continuum fit, or if the models are particularly poor at predicting continuum pixels.\\

We might simply consider what is the best continuum for every possible theoretical spectrum under comparison. When there are thousands of model spectra to compare with, then for each spectrum we can take the ratio of the data and the model, apply some averaging filter (e.g., moving mean or median) and take that as the pseudo-continuum \citep[e.g.,][]{aspcap}. These kinds of approaches work remarkably well, despite the requirement of a (potentially very large) set of sufficiently similar model spectra, which themselves are good descriptions for the data, and the added computational expense.\\

In an ideal scenario the continuum is jointly fit with the stellar parameters. Unfortunately, predicting the emergent spectrum is usually expensive. In Section~\ref{sec:methods} we describe a family of methods to address these problems.
The methods we describe uses non-negative matrix factorisation \citep[NMF;][]{Lee_Seung:2000} to approximate line absorption and telluric transmission. NMF is a linear model to describe large non-negative matrix by two smaller matrices, both with non-negative elements. This non-negativity provides a very useful constraint that is applicable in many areas of astronomy (i.e., where things cannot physically be negative), but NMF has seen relatively little use in astronomy compared to other dimensionality reduction techniques \citep[however see, for example, ][]{Blanton:2007}.
We describe data (Section~\ref{sec:experiments}) and experiments using this model, where the results are presented in Section~\ref{sec:results}. We discuss limitations and potential extensions of our work in Section~\ref{sec:discussion}, before concluding in Section~\ref{sec:conclusions}.\\


\section{Methods}\label{sec:methods}

We will assume a forward model that includes three components: one that represents continuum-normalized stellar lines (e.g., molecular or atomic line absorption); a second to describe telluric lines; and a third to represent the smooth continuum.\footnote{In this work the `continuum' always refers to the joint continuum-instrument response. In many spectrographs these are different, but cannot be disentangled without extra work.} We will require all components to be linear models, which ensures that inference is stable and fast. In practice the linear models we construct seem sufficient to model stellar spectra for the purposes of continuum normalization and handling other nuisances.\\


Here we will describe the method in general before outlining the implementation details. We assume the data are a one-dimensional spectrum with $P$ pixels, where the $i$-th pixel has wavelength $\lambda_i$, flux $y_i$, and flux error $\sigma_{y_i}$ (with $1 \leq i \leq P$). \added{Throughout this work, bold-face symbols denote vectors or matrices (see Table~\ref{tab:nomenclature}).} The forward model for the flux in the $i$-th pixel can be expressed as the element-wise (Hadamard; $\hadamard$) multiplication of what we will describe as the stellar line model $f(\lambda_i; \vecalpha)$, the telluric line model $g(\lambda_i; \vecbeta)$, and the continuum-instrument response model $h(\lambda_i;\vecgamma)$
\begin{align}\label{eq:y}
    y_i &= f(\lambda_i;\vecalpha)\hadamard{}g(\lambda_i;\vecbeta)\hadamard{}h(\lambda_i;\vecgamma) + \mbox{noise}
\end{align}
where the components $f(...)$, $g(...)$, and $h(...)$ are defined below. Throughout this paper we fit in (natural) log-transformed data space $\vec{Y} = \log{\vec{y}}$ with the transformed variance in the diagonal covariance matrix at the $i$-th pixel
\begin{eqnarray}
        \vec{C}_{ii} \approx \frac{\sigma_{y,i}^2}{y_i^2} + \frac{\sigma_{y,i}^4}{2y_i^4}
\end{eqnarray}
\noindent{}which is computed by taking a second-order Taylor series expansion. The log-transformation changes the element-wise product of three model components into the convenient summation:
\begin{align}
    \label{eq:log_y}
    \vec{Y} &= \log{f(\boldsymbol{\lambda}; \vecalpha)} + \log{g(\boldsymbol{\lambda};\vecbeta)} + \log{h(\boldsymbol{\lambda};\vecgamma)} \quad .
\end{align}
We now turn to defining each model component in detail.
The stellar line model $f(\lambda_i;\vecalpha)$ represents the rectified stellar flux at wavelength $\lambda_i$ given parameters $\vecalpha$. We construct the stellar line model $f(\lambda_i;\vecalpha)$ from a set of $N$ continuum-normalized theoretical spectra (each with $D$ fluxes) using non-negative matrix factorisation (NMF). \added{By `continuum-normalized theoretical spectra' we mean the rectified flux provided by spectral synthesis codes, which is the emergent flux divided by the theoretical continuum (i.e., the flux in the absence of line opacity). While the wings of broad lines (e.g., hydrogen) might sometimes be treated as pseudo-continuum in observational work, our theoretical normalization treats all line opacity consistently.}
The theoretical spectra used to construct the stellar line model do not have to have the same wavelength sampling and instrument line spread profile as the data \added{because they can be convolved at inference time, but it is helpful if they are pre-convolved approximately to the target instrument resolution.}\\

\begin{table*}
    \centering
    \caption{Nomenclature}
    \begin{tabular}{cl}
        \hline
        Symbol & Description \\
        \hline
        $P$ & Number of pixels in an observed spectrum\\
        $N$ & Number of continuum-normalized theoretical stellar spectra used in NMF\\
        $D$ & Number of pixels in continuum-normalized theoretical stellar spectra\\
        $K$ & Number of basis components used in the stellar line model\\
        $L$ & Number of basis components used in the telluric line model\\
        $F$ & Number of Fourier basis functions used in the continuum-instrument response model\\
        $\vec{S}$ & Set of continuum-normalized theoretical stellar spectra $[N\,\times\,D]$\\
        $\vecW_\star$ & Non-negative basis weights for stellar line model $[K\,\times\,N]$\\
        $\vec{F}$ & Non-negative basis vectors for stellar line model $[K\,\times\,D]$\\
        $\vecalpha$ & Non-negative weights for the stellar line model found at test time $[K]$\\
        $\vecbeta$ & Non-negative weights for the telluric line model found at test time $[L]$\\
        $\vecW_\earth$ & Non-negative basis weights for telluric line model $[L]$ \\
        $\vecG$ & Non-negative basis vectors for the telluric line model $[K]$ \\
        $\vecbeta$ & Non-negative weights for telluric line model $[L]$ \\
        $\vecH$ & Design matrix for continuum consisting of Fourier bases $[P, F]$\\
        $\vecgamma$ & Log-continuum coefficients found at test time $[F]$ \\
        $\mathbf{\lambda}$ & Wavelengths $[P]$\\ 
        $\mathbf{y}$ & Fluxes $[P]$\\
        $\mathbf{\sigma_y}$ & Uncertainty in fluxes $[P]$\\
        $\mathbf{Y}$ & Log transformed flux $\log{y}$ $[P]$\\
        $\mathbf{C}$ & Diagonal covariance matrix of log-transformed flux $[P\,\times\,P]$\\
        $\hat{\vecalpha}$ & Optimized stellar line model weights $[K]$\\
        $\hat{\vecbeta}$ & Optimized telluric line model weights $[L]$\\
        $\hat{\vecgamma}$ & Optimized continuum-instrument response model coefficients $[F]$\\
        $\hat{\vec{X}}$ & Optimized model parameters (includes $\hat{\vecalpha}$, $\hat{\vecbeta}$, $\hat{\vecgamma}$) $[K + L + F]$\\
        $f(\lambda_i; \vecalpha)$ & Stellar line model \\
        $g(\lambda_i; \vecbeta)$ & Telluric line model \\
        $h(\lambda_i; \vecgamma)$ & Continuum-instrument response model \\
        \hline
    \end{tabular}
    \label{tab:nomenclature}
\end{table*}

\begin{figure*}
    \includegraphics*[width=\textwidth]{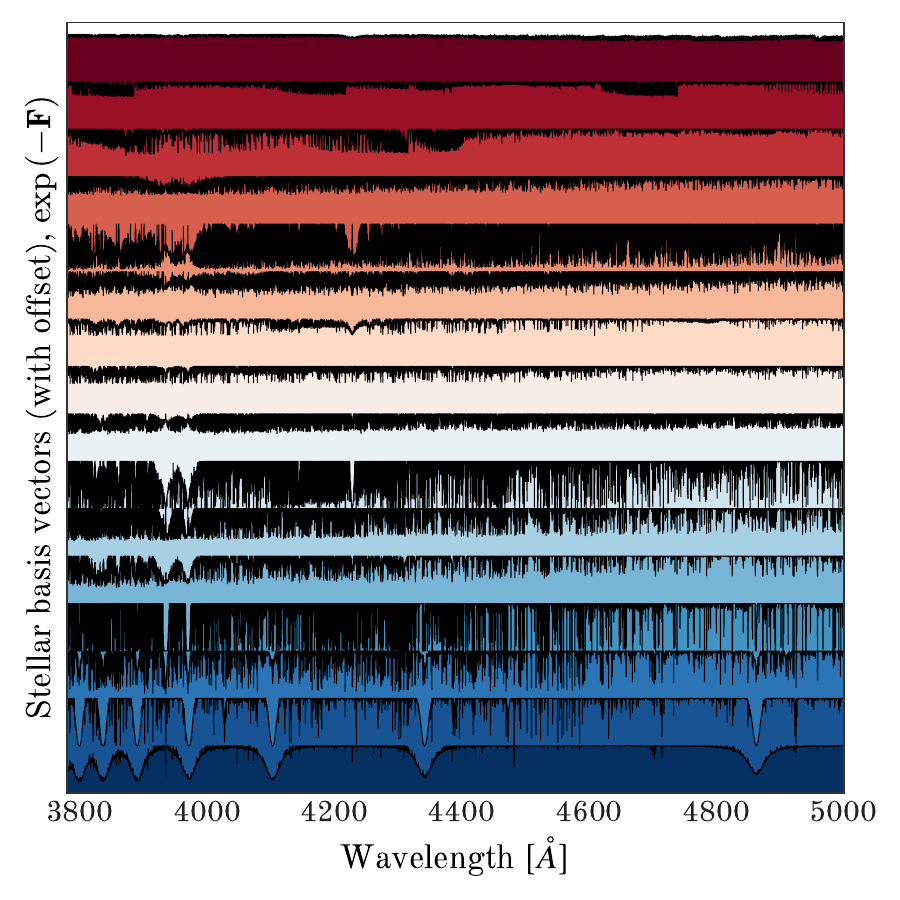}
    \caption{A schematic showing a small wavelength range of the basis vectors computed by non-negative matrix factorisation from a grid of theoretical stellar spectra.\label{fig:schematic}}
\end{figure*}

We refer to our $N \times D$ matrix of continuum-rectified theoretical stellar spectra as $\vec{S}$. This is a dense matrix: there are no entries of exactly zero, with many entries near 1 (no absorption). However, a small transformation to this matrix makes it extremely sparse\added{, which is beneficial because NMF efficiently exploits sparsity. If most pixels have near-zero absorption, then the factorisation can focus on learning the structure of absorption features rather than modelling near-unity continuum values}. Numerous transformations are possible\footnote{Another sparse transformation is $1 - \vec{M}$, but this makes our resulting model non-linear.}, but \added{we chose the negative logarithm transformation because: (1) it converts the multiplicative model (Equation~\ref{eq:y}) into an additive one (Equation~\ref{eq:log_y}); (2) it ensures all transformed values are non-negative (since rectified flux $\leq 1$), satisfying NMF requirements; and (3) it produces sparse matrices because most pixels have flux near unity. We} \removed{for many reasons we chose to} factorise the negative logarithm of $\vec{S}$ into two smaller matrices $\vec{W}_\star$ and $\vec{F}$ such that,
\begin{eqnarray}
    \label{eq:nmf}
    -\log{\vec{S}} \approx \vec{W}_\star\transpose\vec{F}
\end{eqnarray}
where $\vec{W}_\star$\footnote{The star in $\vec{W}_\star$ differentiates it from the basis weights found from factorising telluric spectra $\vec{W}_\earth$.} is a $K \times N$ matrix of \emph{basis weights}, where $K$ is the number of chosen basis vectors, and $\vec{F}$ is a $K \times D$ matrix of NMF \emph{basis vectors}\added{, examples of which are shown in Figure~\ref{fig:schematic}}. The factorisation of $-\log{\vec{S}}$ into $\vec{W}_\star\transpose\vec{F}$ requires that all elements in $-\log{\vec{S}}$, $\vec{W}_\star$, and $\vec{F}$ be non-negative. The number of basis components $K$ should be significantly smaller than both the number of input spectra $N$ and the number of pixels $D$ per spectrum. \added{The choice of $K$ is arbitrary yet easy in practice: too few components yield poor reconstructions, too high would become expensive to factorize. We discuss the choice more in Section~\ref{sec:discussion}, but if we care about getting the continuum correct then we should make the line absorption model as flexible as practical (high $K$). In practice, as few as $K \sim 4$ or 8 works reasonably for low-resolution spectra, with more necessary for higher resolution spectra. In our experiments we select $K$ by trialling multiples of 2 and evaluating the consistency of continuum normalization on a set of random spectra; see Section~\ref{sec:experiments} for specific values used.}\\

The factorisation of $-\log{\vec{S}}$ into $\vec{W}_\star\transpose\vec{F}$ is reasonably fast and easy to compute given existing packages in modern programming languages. In general we found substantial improvements by initializing $\vec{F}$ with non-negative double singular value decomposition, where zeros were filled-in with the average of $-\log{\vec{S}}$: this is the default behaviour in many packages \citep[e.g.,][]{scikit-learn}. We found very similar looking basis vectors if we initialized from random values, but it took more training iterations. We found that factorising $\vec{W}_\star\transpose\vec{F}$ by coordinate descent was substantially faster than using multiplicative updates \citep{Hsieh}. However, the multiplicative update rules are appealing because they are convex \citep{Lee_Seung:2000}, can handle missing or noisy data without requiring many choices \citep{Blanton:2007}, and they can be efficiently iterated in-memory on a graphics card.\\

There are very few requirements of the theoretical spectral grid. There are no requirements on the number of dimensions (e.g., whether or not to include $[\alpha/\mathrm{Fe}]$, $[\mathrm{C/Fe}]$), and no strict requirements (see Section~\ref{sec:discussion}) on spacing between points. The only implicit requirement is that the theoretical spectra should approximately span the range of stars that you intend to fit. This is more of a recommendation than a requirement: in practice we found that a grid trained on theoretical spectra of A- to M-type stars was also sufficiently flexible to model many O-type stars, but there was structured residuals in the wings of hydrogen lines. For these reasons, we chose to include as many theoretical spectra as our memory constraints would allow.\footnote{If the number of spectra exceeds your memory constraints then there are numerous strategies available: you can use memory-mapped arrays, use lower precision float types, compute updates in batches, or simply skip every $n$th spectrum.} \added{In the applications here the training is performed once using hundreds of thousands of spectra for 10,000 steps, which takes approximately 30 minutes on a single CPU core. At inference time, fitting a single HARPS spectrum ($\sim$3$\times$10$^6$ pixels) takes a few core-seconds.}\\

We used the BOSZ spectral grid \citep{Bohlin:2017} to construct basis vectors for a stellar line absorption model. This is a high-resolution ($\mathcal{R} \sim 300{,}000$) finely sampled theoretical grid computed using the \texttt{ATLAS9} code with \project{MARCS} spherical and plane-parallel models atmospheres. \added{The native resolution of the theoretical grid should be at least comparable to the target instrument resolution.} While the grid does include spectra for hotter stars computed with ATLAS atmospheres, we restricted ourselves to MARCS model atmospheres for this work. The grid spans a reasonably large range in stellar parameters, covers the complete wavelength range of the \eso/\harps\ instrument \citep{2003Msngr.114...20M}, and importantly it includes the theoretical continuum for each spectrum. We constructed a sparse kernel to convolve the rectified flux to the \eso/\harps\ spectral resolution of $\mathcal{R} \sim 115{,}000$\added{; this convolution is applied to each spectrum before NMF training, ensuring the basis vectors are learned at the target resolution. In experiments with high rotational broadening (Section~\ref{sec:experiments}) we also apply an rotational convolution to the learned basis vectors}. We truncated the instrument-convolved spectra between 300-700\,nm, and clipped any rectified flux values exceeding 1 \added{(which occur rarely due to numerical precision in the synthesis code or interpolation, not physical emission)} as they violate our NMF requirements. The matrix $\vec{S}$ includes all BOSZ spectra, which we randomly shuffled at every iteration during factorisation. We factorised this matrix into $K = 16$ basis vectors by coordinate descent for {10,000} steps. \added{The choice of $K=16$ stellar basis vectors was arbitrarily determined by starting with $K=2$, testing consistency of continuum-normalization in practice, and doubling $K$ until the performance seemed reasonable. We discuss the choice of $K$ in Section~\ref{sec:discussion}, but in short if we care about modelling the continuum correctly then we should put as much capacity as possible into the other things \citep[e.g., stellar absorption by making $K$ very high;][]{Bickel:1998}.}\\


\noindent{}With the non-negative matrix $\vec{F}$ we can now define the logarithm of stellar line function $\log{f(\lambda_i;\vecalpha)}$. \added{Since $-\log\vec{S} \approx \vec{W}_\star\transpose\vec{F}$ from Equation~\ref{eq:nmf}, we have $\log\vec{S} \approx -\vec{W}_\star\transpose\vec{F}$, and replacing the training weights $\vec{W}_\star$ with inference-time weights $\vecalpha$ yields the log stellar absorption model to be $\vecalpha\transpose\vec{F}$. At inference time there is a transpose necessary to ensure the data are a column vector, leading to}
\begin{eqnarray}
    \log{f(\lambda_i;\vecalpha)} &=& -\vec{F}\transpose\vecalpha \label{eq:log-f} \quad .
\end{eqnarray}
\noindent{}\removed{where $\vecalpha \in [0, \infty)$ is a column vector of $K$ \emph{stellar basis weights} to be solved at inference time.} Here, $\vecalpha \in [0, \infty)$ is analogous to a column vector of \emph{stellar basis weights} in $\vecW_\star$: it represents the $K$ weights needed to reconstruct the rectified stellar flux from the $K$ basis vectors $\vec{F}$. Note that because $\vecalpha$ and $\vec{F}$ are both restricted to have non-negative elements, this severely restricts the flexibility of $f(...)$, leaving $h(...)$ to represent the smooth continuum-instrument response. \added{Specifically, since $\vec{F} \geq 0$ and $\vecalpha \geq 0$, we have $-\vec{F}\transpose\vecalpha \leq 0$, meaning $\log{f} \leq 0$ and thus $f \leq 1$. The stellar model can only produce absorption (flux $\leq 1$), never values exceeding unity. Any observed flux above the stellar model prediction must therefore be captured by the continuum component $h(...)$.}\\

Telluric lines can be included in a similar way to stellar lines. We used a grid of model telluric transmission spectra computed for the \texttt{PypeIt} project \citep{pypeit}. These spectra are computed using the Line-by-Line Radiative transfer Model \citep{lblrm} for an array of observatory locations (latitude, longitude) and atmospheric temperatures, pressures, humidity, and airmass. The original grid of telluric transmission spectra was not available to us. Instead, we used a representation of this grid that is shipped with \texttt{PypeIt}, where the grid is compressed using Principal Component Analysis (PCA) and an inverse hyperbolic sine transformation. We reconstructed the original grid from these approximations, and factorised the negative log of theoretical telluric absorption
\begin{eqnarray}
    -\log\vec{M} \approx \vec{W}\transpose_\earth\vec{G}
\end{eqnarray}
\noindent{}into two-non negative matrices $\vec{W}\transpose_\earth$ and $\vec{G}$ using $L = 4$ basis vectors\added{, the dimensionality of the PCA-compressed grid}. We made {10,000} iterations using coordinate descent. With the non-negative matrix $\vec{G}$\added{, like Equation~\ref{eq:log-f},} the function $\log{g\left(\lambda_i;\vecbeta\right)}$ \added{becomes}\removed{is simply}
\begin{eqnarray}
    \log{g(\lambda_i;\vecbeta)} = -\vec{G}\transpose\vecbeta \label{eq:log-g}
\end{eqnarray}
where $\vecbeta \in [0, \infty)$ is column vector of $L$ \emph{telluric basis weights} in to be solved at inference time.
Like $\vecalpha$, $\vecbeta$ is analogous to a column \added{vector} in $\vec{W}_\earth$: it represents the $L$ basis weights needed to reconstruct the telluric transmission.\\

There are many suitable choices for the logarithm of the continuum-instrument response model $\log{h(\lambda_i;\vecgamma)}$. Here we chose a Fourier basis of \added{$F$} sine and cosine functions because it is a linear representation, and is sufficiently flexible for modelling the joint continuum-instrument response across a variety of spectrographs. The component $\log{h(\lambda_i;\vecgamma)}$ is expressed compactly as
\begin{align}
    \log{h(\lambda_i;\vecgamma)} = \vec{H}\vecgamma \label{eq:log-h}
\end{align}
where $\vec{H}$ is a design matrix where the elements of the $j$th column are, 
\begin{align}
    \vec{H}_{j}(\lambda_i) & = \left\{\begin{array}{cl}\displaystyle\cos\left(\frac{\pi\,[j-1]}{P}\,\lambda_i\right) & \mbox{for $j$ odd} \\[3ex]
    \displaystyle\sin\left(\frac{\pi\,j}{P}\,\lambda_i\right) & \mbox{for $j$ even}\end{array}\right. ~,
\end{align}
\noindent{}where $P$ is a length scale which we set to be twice the length of the wavelength range to fit (e.g., twice the peak-to-peak range of an echelle order). 
Throughout this paper we will describe $\vecgamma$ as the sine-and-cosine \emph{coefficients}.\\

\noindent{}With $f(\lambda_i;\vecalpha)$, $g(\lambda_i;\vecbeta)$ and $h(\lambda_i;\vecgamma)$ now defined, we can expand the forward model in Equation \ref{eq:log_y} for the log-transformed data $\vec{Y}$ \added{using Equations~\ref{eq:log-f}, \ref{eq:log-g}, and \ref{eq:log-h}}:
\begin{equation}
    \vec{Y} = -\vec{F}\transpose\vecalpha - \vec{G}\transpose\vecbeta + \vec{H}\vecgamma
\end{equation}
\noindent{}which we can compactly write as
\begin{equation}
    \vec{Y} = \vec{A}\vec{X}
\end{equation}
where the \removed{parameters and design matrices are stacked to construct}
\added{design matrices are stacked to construct} $\vec{A}$ and \added{the unknown parameters are stacked into} $\vec{X}$:
\begin{eqnarray}
    \vec{A} = \begin{bmatrix}-\vec{F}\transpose\\-\vec{G}\transpose\\\vec{H}\end{bmatrix}
    \quad \mbox{and} \quad
    \vec{X} = \begin{bmatrix}\vecalpha\\\vecbeta\\\vecgamma\end{bmatrix} \quad .
\end{eqnarray}

Given some log-transformed observed flux $\vec{Y}$ and associated covariance matrix $\vec{C}$, this reduces to a weighted least-squares problem. This could be solved directly by linear algebra,
\begin{eqnarray}
    \vec{\hat{X}} = (\vec{A}\transpose\vec{C}^{-1}\vec{A})^{-1}\vec{A}\transpose\vec{C}^{-1}\vec{Y}
\end{eqnarray}
\noindent{}but this solution is not subject to the constraints $\vecalpha \geq 0$ and $\vecbeta \geq 0$. There are a few ways to solve this system subject to the boundary constraints, but the choice is not very important because the problem is convex. For simplicity we use the truncated reflective trust region algorithm where we stop optimizing once the relative improvement has reached machine precision. \added{We denote the optimised parameters with hats (e.g., $\boldsymbol{\hat{\alpha}}$, $\boldsymbol{\hat{\beta}}$, $\boldsymbol{\hat{\gamma}}$) to distinguish them from the general parameters.}
After solving for $\vec{\hat{X}}$, the (un-transformed) quantities can be easily computed:
\begin{eqnarray}
    \mathrm{rectified~stellar~flux}, \quad &f& = \exp{\left(-\vec{F}\transpose\boldsymbol{\hat{\alpha}}\right)} \\
    \mathrm{telluric~transmission},\quad  &g& = \exp{\left(-\vec{G}\transpose\boldsymbol{\hat{\beta}}\right)} \\
    \mathrm{continuum}, \quad &h& = \exp{\left(\vec{H}\boldsymbol{\hat{\gamma}}\right)}
\end{eqnarray}

\section{Experiments} \label{sec:experiments}

The model we describe has most value in situations where there are lots of data, and many nuisance parameters. One example is high-resolution echelle spectra where there are different continuum parameters per order, and the large number of data points (e.g., $\sim10^6$~pixels/spectrum) can make even the fastest non-linear interpolators (or spectrum emulators) untenably slow.\\

We assess the model using two experiments with real data. In Experiment 1 we test the model on spectra across a wide range of spectral types so that we can evaluate whether a constrained linear absorption model is an effective representation for typical stars. In Experiment 2 we fit the model to repeat observations of the same star, taken across many different S/N ratios, to test how consistent the rectified spectra are as a function of S/N ratio.\\

We queried a catalog of \eso/\harps\ \citep{2003Msngr.114...20M} observations to select a high-resolution echelle spectrum of each spectral type from M- to B-type for Experiment 1. This selection is not intended to be complete, but is meant to demonstrate the effectiveness of a single linear model for fitting stars across a range of stellar parameters and evolutionary states. For each example star we downloaded a random exposure. No explicit cut was made to ensure a minimum S/N ratio, but in practice most exposures are of high quality. \added{Table~\ref{tab:sample} summarises the stars used and their approximate S/N ratios.} The B- and A-type example stars show significant rotational broadening. For these stars, we adopt the $v\sin{i}$ values reported by \citet{Barbieri:2023} and convolve the \added{already-learned} stellar basis vectors before inference. In Section~\ref{sec:discussion} we discuss situations when radial velocity or rotational broadening measurements are not available.\\

\added{
\begin{table}
    \centering
    \caption{Summary of example stars used in Experiment 1.}
    \begin{tabular}{lccc}
        \hline
        Star & Spectral Type & Approx.\ S/N & Notes \\
        \hline
        HD 45418  & B4 & 150 & Blaze-corrected \\
        HD 186543 & A7 & 200 & Blaze-corrected \\
        HD 222595 & F6 & 180 & -- \\
        HD 20794  & G8 & 250 & -- \\
        HD 203050 & K2 & 120 & Red giant \\
        HD 85512  & K6 & 100 & -- \\
        HD 95735  & M1 & 80 & -- \\
        \hline
    \end{tabular}
    \label{tab:sample}
\end{table}
}

\begin{sidewaysfigure}
    \includegraphics[width=\textwidth]{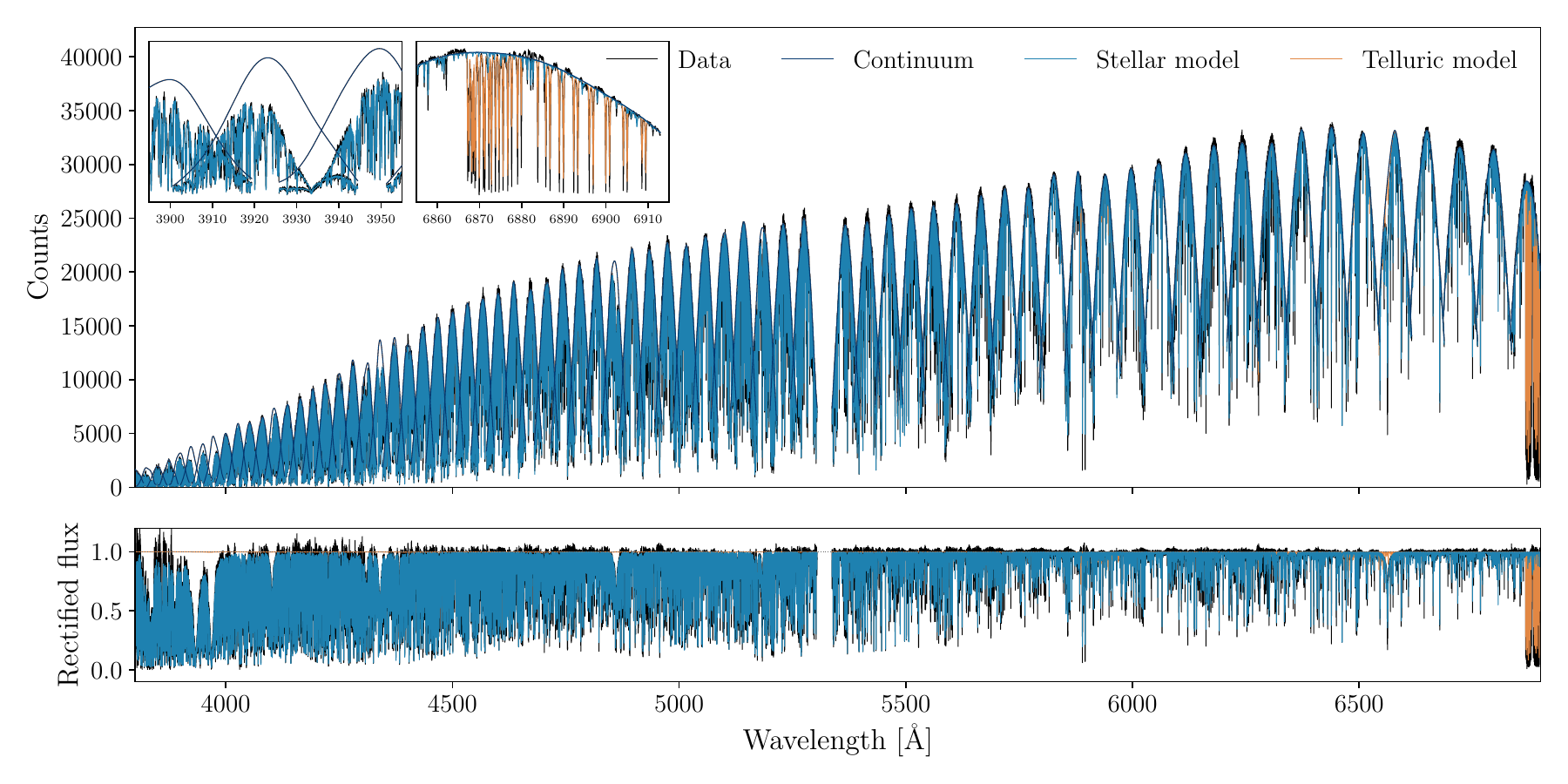}
    \caption{An example fit to a single \eso/\harps\ spectrum of HD~222595 (black), showing the continuum fits per echelle order \removed{(dotted)}, the stellar absorption model (\removed{red}\added{blue}) and the telluric transmission model (\removed{blue}\added{orange}), all fit simultaneously with a constrained linear absorption model. Near the strong Ca H and K lines (see inset) it is clear that there are \emph{no} `continuum pixels' in those echelle orders, yet the model provides an excellent estimate of the true continuum. \added{The telluric model used is imperfect (see inset), but sufficient to capture most telluric absorption.}\label{fig:example-echelle}}
\end{sidewaysfigure}

The final product of the \harps\ data reduction pipeline is a one-dimensional unrectified spectrum, resampled onto a single wavelength array where all echelle orders have been stitched together. Because we are interested in fitting the continuum on a per-order basis, we instead use the \texttt{e2ds} data products. These are extracted one-dimensional unrectified spectra of each echelle order, with the wavelength solution represented by a low order polynomial. In early experiments we fit the spectra in the \texttt{e2ds} data products directly, an example of which can be seen in Figure~\ref{fig:example-echelle}. However, for HD 186543 (an A7 star) and HD 45418 (B4) we noticed systematic residuals that could be explained by a 2D artefact in the image plane (e.g., a ghost or reflection). While we expect that flat-fielding should have already removed these kinds of artefacts, it did prompt us to explore using the nightly blaze calibration files. Using the calibration blaze files mitigated much of the visible structure, providing spectra with a simpler, flatter shape. For this reason, for HD 186543 and HD 45418 we fit blaze-corrected echelle orders. \added{We leave these examples in to demonstrate the applicability of our method to both blaze-corrected and blaze-uncorrected spectra.}\\

When fitting each exposure we took the computed Barycentric Earth radial velocity and the radial velocity measured by the \harps\ data reduction pipeline to shift the stellar absorption basis vectors $\vec{F}$ to the observed frame. In situations where no radial velocity was available from the \harps\ data reduction pipeline, we adopted the radial velocity reported by \citet{Trifonov:2020}. Each exposure has $\sim3\times10^6$ pixels across 72 echelle orders, which we model with 16 stellar basis vectors and 4 telluric vectors \added{(see earlier justification)}, and {9} Fourier modes per echelle order\added{, which seemed sufficient flexibility to capture the continuum shape}, totalling {668} model parameters per spectrum.\\

We used the same model setup for \removed{the} Experiment 2\added{, where we collected many spectra of the same star at many (e.g., between 10 to 200) S/N ratios}. \removed{In this experiment we collected many spectra taken of the same star at low and high (e.g., 10 to 200) S/N ratios.} The lower bound on S/N ratio might be unrealistic since spectroscopists prefer measuring stellar parameters and chemical abundances from higher quality spectra, but the bound is necessary to evaluate performance at low S/N ratios. The \eso/\harps\ catalog revealed that not many stars meet our repeated S/N sampling requirement. When sorted by number of exposures in descending order, $\alpha$-Centauri A was the first star that met our requirements in repeated S/N sampling. We constructed bins of S/N from 0 to 200 (with a bin width of 1)\removed{in steps of 10}, and selected a random $\alpha$-Centauri A spectrum for every S/N ratio bin interval to provide an even sampling of spectra at low and high S/N ratios. We fit our model to each spectrum, treating it as totally independent from all others.\\

\begin{figure*}
    \includegraphics[width=\textwidth, height=\textwidth]{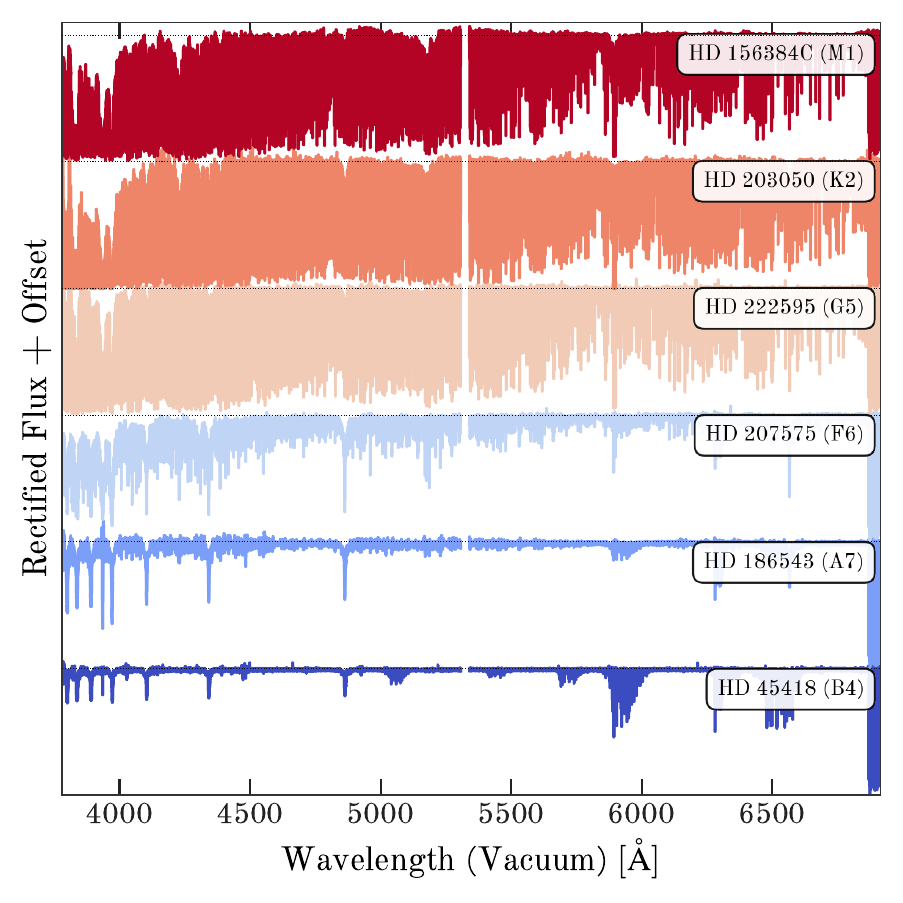}
    \caption{Continuum-rectified \harps\ spectra of example spectral types from M to B. The model used here includes 16 basis vectors for stellar absorption, and 4 basis vectors for telluric transmission.}\label{fig:harps-examples}
\end{figure*}

\section{Results} \label{sec:results}

Experiment 1 involves fitting stars of a variety of spectral types. The rectified spectra for selected example stars in this experiment are shown in Figure~\ref{fig:harps-examples}, where stars are ordered and coloured by their spectral type listed in \project{SIMBAD}. From these rectified spectra it is clear that the model performs well on a variety of stellar types and evolutionary stages. HD 45418, a B4-type star, is the hottest star in our sample  \citep[$T_\mathrm{eff} = 16{,}750\,K$ according to ][]{2015MNRAS.454...28M}, and far exceeds the upper temperature bound (8,000\,K) of the synthetic grid that we factorised using NMF. The fit to HD 45418 is quite good, despite this star being well outside the model grid. All other example stars shown in Figure~\ref{fig:harps-examples} have literature stellar parameters that place them well within the bounds of the factorised model grid. Most stars observed with \eso/\harps\ are on the main-sequence, since they are more frequently subject to radial velocity exoplanet searches than evolved stars. HD 203050 (K2) is the only red giant branch in our sample. This is not obvious from Figure~\ref{fig:harps-examples}, suggesting the model does just as well on red giant branch stars as it does on main-sequence stars. This has been our experience in other works using this method on SDSS-V optical and near-infrared spectra (Casey et al., in prep).\\

The cooler stars in the example set have almost no so-called `continuum pixels' blueward of 4000\,\AA. This is expected; the continuum estimate provided by the model matches theoretical rectified spectra. It illustrates that the continuum estimate that we fit simultaneously with stellar absorption does provide a closer estimate to the true continuum rather than a pseudo-continuum. \added{However, this region presents a greater challenge for continuum normalization, as the high density of absorption features makes it difficult to identify true continuum points.}\\

There is good agreement between the model fits and data for example stars of different spectral types. In detail there are mis-matches in line strength, or mis-matches in the form of missing lines (in either the model or the data). We discuss model mismatches in Section~\ref{sec:discussion}, but we can surmise that model mis-matches are a result of missing or inaccurate physics in the spectral synthesis code or its inputs (e.g., model atmospheres, line lists, opacities), or due to the poor linear approximation we make for the stellar spectra, or both. In any case, the model appears to be sufficiently accurate to describe the continuum without substantial bias, so it is not a major concern for our purposes. \added{If the theoretical models were perfect descriptions of stellar spectra, and our NMF approximation was also perfect, then any pixels exceeding the dashed line in Figure~\ref{fig:harps-examples} should be consistent with photon noise and emission lines. However, neither is true. This motivates that we should add flexibility to our stellar absorption model by increasing $K$ or building the stellar absorption model from observed spectral templates.}\\

A separate reason for model-data disagreement comes from our use of the telluric transmission model. The telluric models available to us are of lower spectral resolution ($\mathcal{R} \sim 25{,}000$) than \harps. This is apparent when closely comparing the fitted model with the data (e.g., if one could zoom in finely to Figure~\ref{fig:example-echelle}), but it is otherwise not obvious. The model telluric transmission predicts features that are clearly broader than the data, but the telluric basis vectors are a sufficiently good description for the data that the telluric absorption does not get captured by the continuum basis vectors and adversely affect the fit. \added{In other words, the telluric features describe narrow atmospheric features while the continuum basis vectors describe smooth functions.} \added{Higher-resolution telluric templates (e.g., ESO \texttt{SkyCalc}) could improve the fit in telluric-contaminated regions; our choice of the \texttt{PypeIt} grid was driven by convenience rather than optimality. Our approach can readily accommodate any telluric grid.}\\

\begin{figure}
    \includegraphics*[width=\textwidth, height=3in]{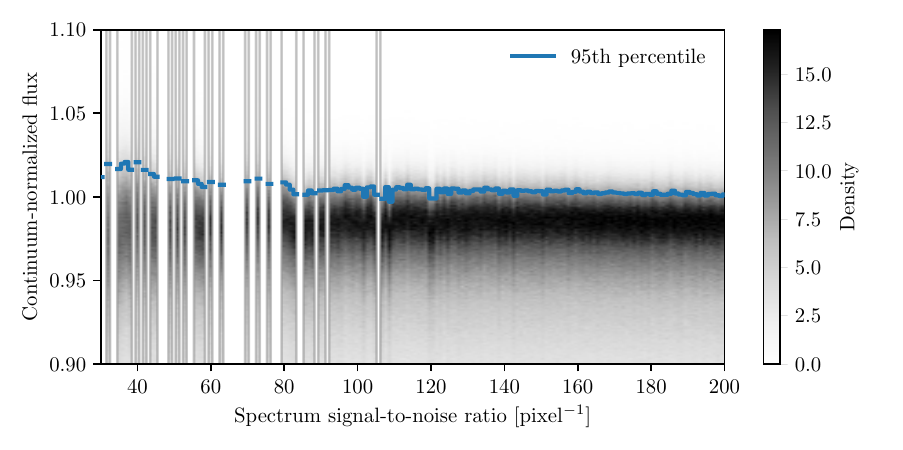}
    \caption{Distribution (shaded) of rectified flux values of $\alpha$-Centauri A as a function of S/N ratio, with every spectrum fit independently. The shaded region line represents distribution of flux values for each spectrum (e.g., each vertical distribution represents all $\sim3\times10^6$ pixels in a single spectrum). The blue line indicates the 95th percentile, which has a 1-$\sigma$ spread of 0.46\% for spectra with S/N $>$ 30, and 0.22\% for everything with S/N $>$ 100.\label{fig:repeatability-with-snr}}
\end{figure}

Experiment 2 tests how consistently we can predict continuum for the same star as a function of S/N ratio. Given an imperfect model (or no model) for stellar spectra, noisy signals can mimic stellar absorption, which causes the continuum to become systematically biased at low S/N ratios. We fit repeat spectra of $\alpha$-Centauri A, which are sampled uniformly in S/N bins from 10 to 200 pixel$^{-1}$. This experiment forces us to select a summary statistic to evaluate the consistency of our continuum rectification. We adopt the 95th percentile in rectified flux as a statistic to compare to normalization methods described in the literature, but we prefer to view how the \emph{distribution} of normalized flux values change as a function of S/N ratio. Both properties are shown in Figure~\ref{fig:repeatability-with-snr}, where the line indicates the 95th percentile of normalized flux, and the shaded area shows the distribution of normalized flux values. Above S/N $\sim$ 30, both indicators are extremely stable and do not vary significantly with S/N ratio. This is notable, as most spectra acquired for the purpose of stellar parameters and chemical abundances will aim for S/N $>$ 30, but spectra below S/N $<$ 50 are routinely described as being problematic to rectify consistently \citep{Cretignier:2020}. From this figure it highlights that we can achieve consistent continuum normalization whether the star has S/N $\sim$ 30 or 300. Notably, as stated earlier, in this experiment we did not use blaze-corrected spectra.\\

We can make quantitative statements on the consistency of continuum rectification for $\alpha$-Centauri A. Taking the 95th percentile of normalized flux (i.e., observed flux / estimated continuum) as a proxy statistic for the normalization consistency, we find that above S/N $> 100$, the standard deviation of the 95th percentile of normalized flux is 0.22\%, which is equivalent a spread in normalized flux of 0.0022. This variation is smaller than the typical flux errors, and only increases marginally when we include lower quality spectra. Taking all spectra with S/N $>$ 50, we find the scatter in the 95th percentile of normalized flux to be 0.35\%. With S/N $>$ 30, this rises marginally to 0.46\%. In all scenarios the estimated error in continuum is less than the typical flux error.\\


\section{Discussion}\label{sec:discussion}

We have shown that a constrained linear model is a good choice for modeling stellar spectra, which enables precise estimates of the continuum for a range of stellar types. Representing rectified stellar flux using NMF instead of a some other linear representation is particularly advantageous. Non-negativity ensures that stellar basis vectors are strictly additive, and the negative log-transformation we use restricts both the rectified stellar flux and telluric transmission to be between 0 and 1. Any data values outside this range \emph{must} either be captured by the continuum basis vectors, or be noise. In some sense thestellar line basis and the continuum basis are approximately orthogonal.\\

NMF is less frequently used in astronomical data analysis than other linear representations. A more common linear representation for this kind of problem is Principal Component Analysis \citep[PCA;][]{pca}, which yields an orthogonal decomposition that maximally explains the variance in the data. PCA is made possible by orthogonality, whereas NMF is made possible by the non-negativity constraint. While PCA can capture more variance than NMF for the same data set and number of components, in PCA there are no bounds on the eigenvalues (basis weights). For this reason, if we used PCA to represent stellar spectra then it would make it nearly impossible to determine the continuum. We could model the data without having any continuum basis vectors, simply by finding the first eigenvalue to be a very large positive value, the second eigenvalue to be a large negative value, the third a slightly lower (but still large) positive value, and so on. This can easily model flux values much larger than 1. The situation does not get much better if we include basis vectors for the continuum, because much of that smooth structure can be captured by the PCA spectral model. \added{One might argue that truncating PCA to a limited number of components would ameliorate this issue, but the fundamental problem remains: PCA basis vectors and weights can take any sign, so even with few components the model can produce unphysical flux values. While both PCA and NMF require choosing the number of components (an admittedly arbitrary choice), NMF's non-negativity provides a physical constraint that PCA lacks --- it guarantees the stellar model produces only absorption ($f \leq 1$), regardless of how many components are used.}\\

Another \removed{somewhat unexpected} advantage of using NMF is that the basis vectors themselves appear to be \added{qualitatively} \removed{highly} interpretable. For example, one basis vector has only strong hydrogen lines in it. Another only shows molecular CO features that only appear in M-type stars. Another two basis vectors are similar in that they have a forrest of atomic lines, but they vary in the broadness of particularly strong lines (e.g., where one is consistent with pressure broadening in a main-sequence star). General wisdom when working with compressed representations is that the basis vectors should generally be treated as latent (unobservable) vectors, and there is risk associated with trying to physically interpret them. We avoid \added{detailed} physical interpretability here, but in subsequent work we discuss how the basis weights and vectors can be interpreted (Casey et al., in prep). Here we simply reiterate that the stellar parameters used to construct the grid are not used by the NMF process, and the spectra are randomly shuffled during each iteration of factorisation, so there is no possible `leakage' of information about effective temperature, surface gravity, metallicity, or other properties. Nevertheless, the resultant basis vectors have absorption features that are strongly correlated with these properties. \\

\added{Our model occupies a middle ground between purely data-driven methods and traditional forward modelling. The NMF basis is derived from theoretical spectra, encoding physical priors about stellar absorption, but at inference time the basis weights are determined by the data alone without imposing a mapping to stellar parameters.} \removed{Chosing not to interpret the basis vectors or weights has specific advantages.} For example, if we were to build some mapping from stellar parameters to basis weights, then one could construct a forward model for the data where the model parameters are the continuum coefficients and the stellar parameters (e.g., $T_\mathrm{eff}$, $\log{g}$), instead of continuum coefficients and basis weights. That model would be far less flexible than the model we present here. For example, our model is capable of providing excellent fits to the data, even if the underlying theoretical spectra are inaccurate. A simple example is hydrogen lines. \citet{Wheeler:2024} showed that the hydrogen lines in the infrared are incorrectly computed by many spectral synthesis codes. When we use existing grids with this problem (e.g., like those used in \texttt{FERRE}, as part of \texttt{ASPCAP}), if we had a model that was mapping from stellar parameters to basis weights, then even if the hydrogen lines are in their own basis vector, we would find a biased estimate of effective temperature, or a worse fit to the data, or both. Instead, by not restricting ourselves to a mapping between stellar parameters to basis weights, we retain flexibility to make excellent fits to the data. \added{We caution, however, that an excellent fit does not guarantee accurate stellar parameters --- our model is designed for continuum normalization, not parameter inference. The flexibility that enables good fits could also mask model deficiencies; users should not interpret small residuals as validation of the underlying physics.}\\

Our model is limited by how well it predicts data in two different ways. The first is in how well the models of stellar atmospheres, line lists, and spectral synthesis codes can match real data. There is very little we can do about that error mis-match here. \added{However, one advantage from this factorization is that we learn absorption \emph{vectors}, and the weights we fit at test time do not need to represent a set of weights that exist in the training set. For example, if TiO absorption was under-predicted by spectral synthesis codes, we could simply infer a slightly higher basis weight for the basis vector(s) that describe TiO absorption. A related concern is if there are absorption features that are simply not modelled at all by spectral synthesis codes. Such a scenario might be unmodelled molecular features that get absorbed by the continuum component, biasing the normalization. The non-negativity constraint limits this effect to a degree: the continuum can only explain flux \emph{above} what the stellar model predicts, not flux deficits.} The second form of error arises in our factorisation of the grid. We could add more basis vectors, train for longer, even include estimated per-pixel model uncertainties, but the reconstruction will never be perfect. A related minor deficiency in our model choice is that it cannot model emission lines in stellar spectra: the model can only predict rectified flux values between zero and one. This is a minor problem because emission lines are rare, only occurring in the strongest of lines for chromospherically active stars, and because the emission lines only affect a small number of pixels, they contribute very little to the overall $\chi^2$.\\

The speed of our model makes it particularly advantageous for evaluating on large data sets. If we assume that the factorisation of stellar spectra introduces a small error relative to the usual mis-match between models and data, then a constrained linear model represents a promising avenue for identifying regions where models and observations disagree the most. \added{Unlike traditional forward modelling where fitting a single spectrum can take minutes to hours, our approach fits a spectrum in seconds, enabling systematic analysis of millions of stars.} The simplest path for doing this might be to look at the total squared residuals as a function of pixel across many stars, which would immediately indicate the areas where the model does not have capacity to predict the data. Next steps might be to improve modelling for that region (e.g., through better line lists) or might be to iteratively update the basis vectors in a data-driven way so that they better match the data. The model is trained from synthetic spectra, so it is already a close approximation to what stellar spectra look like, and doing data-driven updates would be an excellent way to handle severe data-model mis-matches (e.g., large molecular features which are present in data but we lack line lists to accuratley model). There is a lot of flexibility when updating the basis vectors using multiplicative update rules \citep{Blanton:2007}: we can restrict ourselves to only allowing updates on particular pixels (where there is sharp disagreement), or we can restrict ourselves to only updating a single basis vector. \\

There are not many literature metrics to choose from when evaluating performance of continuum normalization. We have qualitatively shown the distribution of rectified flux values for $\alpha$-Centauri A, and quantitatively estimated the precision from the scatter in the 95th percentile of rectified flux values. These statistics are computed using all wavelengths, but it is well-known that continuum is easier to model at redder optical wavelengths because there are fewer absorption features. One excellent examination of continuum normalization \removed{consistency} is that of \citet{Cretignier:2020}, where they report \removed{consistent} \added{\texttt{RASSINE} provides} continuum normalization \removed{with \texttt{RASSINE}} \added{accurate} at the level of 2\% based on applications to extremely high quality ($S/N > 1000$) \eso/\harps\ spectra of $\alpha$-Centauri B stacked over consecutive nights, and 0.29\% for a comparably high quality spectrum of the Sun. \added{We note that these metrics from \citet{Cretignier:2020} represent \emph{accuracy} (comparison to theoretical models at anchor points), not \emph{consistency} across different observations. Since any systematic offset between model and \texttt{RASSINE}-derived continuum would be constant across time, the reproducibility of \texttt{RASSINE} may be better than these accuracy figures suggest.} Their statistic is defined as the 2$\sigma$ of $1-f_i$, where $f_i$ is the $i$th normalized flux value at the anchor point locations used to build the continuum. The anchor points are chosen by the user, which makes for a less direct comparison with our statistic\added{, as we evaluate consistency across different observations rather than accuracy against models}. \removed{but for these purposes we will assume that the continuum anchor points are perfectly chosen and truely represent the continuum in the star. Given these assumptions, for comparison purposes we take that an expert user of \texttt{RASSINE} \citep{Cretignier:2020} can achieve 1\% consistency in continuum estimates for S/N $> 1000$ spectra of $\alpha$-Centauri B, and 0.15\% consitency for comparable Solar spectra.} If \citet{Cretignier:2020} restrict their statistic to wavelengths redder than 4500\,\AA, their \added{accuracy} \removed{precision} for $\alpha$-Centauri B improves to 0.6\%. \added{Our metric measures a different quantity --- the reproducibility of the 95th percentile of rectified flux across independent observations --- but the comparison illustrates that our method achieves sub-percent consistency even at S/N ratios an order of magnitude lower than the stacked \texttt{RASSINE} spectra.}\\



Arguments on the consistency of continuum normalization are precision-based arguments, which says nothing about accuracy. That is unfortunate for us because any pseudo-continuum method explicitly only needs to make arguments about precision: a pseudo-continuum is not intended to be accurate! In our case, our model is built to try and estimate the true continuum as best as possible by properly accounting for the stellar absorption. While this does likely lead to a more accurate estimate of the continuum, measuring that accuracy is very hard. Getting the continuum absolutely correct requires a stellar spectrum of a flux-calibrated instrument, and it requires that we have good models of stellar absorption. Some instruments deliver flux-calibrated spectra, but we rarely believe that we have good models of stellar absorption for M-type stars. Here, we think we do have good models of stellar absorption for M-type stars, but there are known deficiencies: the CaH and TiO bands are poorly modelled in most synthetic spectral grids, including the one we used. 
Put simply, measuring accuracy of continuum is hard, and we don't try it\added{. We argue our estimate is preferable to a pseudo-continuum for several reasons: (1) it is physically motivated, using theoretical spectra that encode knowledge of where absorption features occur; (2) it is reproducible across S/N ratios without parameter tuning; and (3) in regions with no true continuum pixels (e.g., blue wavelengths for cool stars), our method still provides a meaningful estimate while pseudo-continuum methods must either extrapolate or fail. However, we acknowledge that in regions where models are poor (e.g., molecular bands), a well-chosen pseudo-continuum from an expert practitioner might be more accurate than our model-based estimate}.\removed{, but we think our continuum estimate is better than a pseudo-continum.}\\

We used the radial velocities reported by the \eso/\harps\ pipeline to shift stellar basis vectors to the observed frame. In cases of extreme rotational broadening, we also used literature measures of $v\sin{i}$. We took both of these quantities as truth, but in fact we could fit for them simultaneously. A radial velocity shift and a convolution are both linear operators, in that they can be incorporated in our model without forcing the model to become strongly non-linear. We would have to change how we solve the system: we could no longer solve the system directly by unconstrained least-squares, or with the truncated reflective region algorithm, but we would have to solve it using linear operators. Alternatively, given the speed of our model, one could imagine a grid in radial velocities (e.g., in $\pm1\,000\,\mathrm{km}\,\mathrm{s}^{-1}$), where at each radial velocity we solve for $\vec{\hat{X}}$ and then take the radial velocity with the lowest $\chi^2$. This is a robust method. The rotational broadening could be estimated from the cross-correlation in the same way. For these reasons, the method we describe here is readily extensible to situations where the radial velocity and the rotational broadening is not known.\\

\section{Conclusions} \label{sec:conclusions}

We introduce a novel approach to modelling stellar spectra as a linear combination of highly constrained absorption vectors, which we show is sufficient for simultaneously fitting stellar lines, telluric transmission, and the joint continuum-instrument response. \added{A one-time training step factorises a grid of theoretical spectra to produce the basis vectors, with the number of components $K$ as the primary hyperparameter.} A constrained linear absorption model has many advantageous properties over alternative data-driven methods or machine learning techniques used in stellar spectroscopy. The model structure restricts basis vectors to be strictly additive, ensuring separability between the basis vectors for absorption and continuum\added{, with the non-negativity constraint guaranteeing that the rectified stellar flux is restricted to $f \leq 1$}. The model requires no initial guess or prior of the stellar parameters \added{at inference time}, and the linearity ensures that inference is robust, stable, and fast. \added{We demonstrate consistency in continuum normalization at the 0.2\% level for S/N $>$ 100 spectra and better than 0.5\% at S/N $\sim$ 30.} An extension of this work could also solve for other linear nuisances (e.g., radial velocity, rotational broadening). \removed{Finally, an unexpected result of this work is that the basis vectors found from factorising model rectified spectra appear to be highly interpretable, which is promising for their use for interpretable constrained linear absorption models.} \added{The basis vectors appear qualitatively interpretable, which we explore further in subsequent work (Casey et al., in prep).}\\




\paragraph{Software}
\texttt{numpy} \citep{numpy};
\texttt{matplotlib} \citep{matplotlib};
\texttt{scipy} \citep{scipy};
\texttt{scikit-learn} \citep{scikit-learn}.

\paragraph{Acknowledgements}
We thank the anonymous referee for a constructive and detailed report that improved the quality of this manuscript.
A.~R.~C. thanks
    Rory Smith (Monash University),
    Guy Stringfellow (Boulder),
    Julianne Dalcanton (Flatiron Institute), and
    Michael Blanton (NYU).
A.~R.~C. thanks New York University for their hospitality.

\bibliographystyle{aasjournal}
\bibliography{bibliography}

\end{document}